\documentclass[nonacm,acmlarge,screen]{acmart}

\AtBeginDocument{%
  \providecommand\BibTeX{{%
    \normalfont B\kern-0.5em{\scshape i\kern-0.25em b}\kern-0.8em\TeX}}}
\copyrightyear{2023}
\acmYear{2023}
\setcopyright{cc}
\setcctype{by-nc-sa}
\acmConference[CHI '23 Workshop]{CHI '23 Designing Technology and Policy Simultaneously Workshop}{April 23,
2023}{Hamburg, Germany}
\acmDOI{}
\acmISBN{}

\begin{document}
\fancyhead[RO]{\fontfamily{LinuxBiolinumT-TLF}\fontsize{8}{10}\selectfont CHI '23 Workshop, April 23, 2023, Hamburg, Germany}

\fancyhead[LE]{\fontfamily{LinuxBiolinumT-TLF}\fontsize{8}{10}\selectfont Designing Technology and Policy Simultaneously Workshop}


\title{Promoting Bright Patterns}
\subtitle{\url{https://brightpatterns.org/}}

\author{Hauke Sandhaus}
\email{hgs52@cornell.edu}
\orcid{0000-0002-4169-0197}
\affiliation{%
  \institution{Cornell University, Cornell Tech}
  \streetaddress{2 West Loop Rd}
  \city{New York}
  \state{New York}
  \country{USA}
  \postcode{10044}
}
\begin{abstract} 
User experience designers are facing increasing scrutiny and criticism for creating harmful technologies, leading to a pushback against unethical design practices. While clear-cut harmful practices such as dark patterns have received attention, trends towards automation, personalization, and recommendation present more ambiguous ethical challenges. To address potential harm in these “gray” instances, we propose the concept of “bright patterns” – persuasive design solutions that prioritize user goals and well-being over their desires and business objectives. The ambition of this paper is threefold: to define the term “bright patterns”, to provide examples of such patterns, and to advocate for the adoption of bright patterns through policymaking.
\end{abstract}

\begin{teaserfigure}
\begin{centering}
  \includegraphics[width=0.8\textwidth]{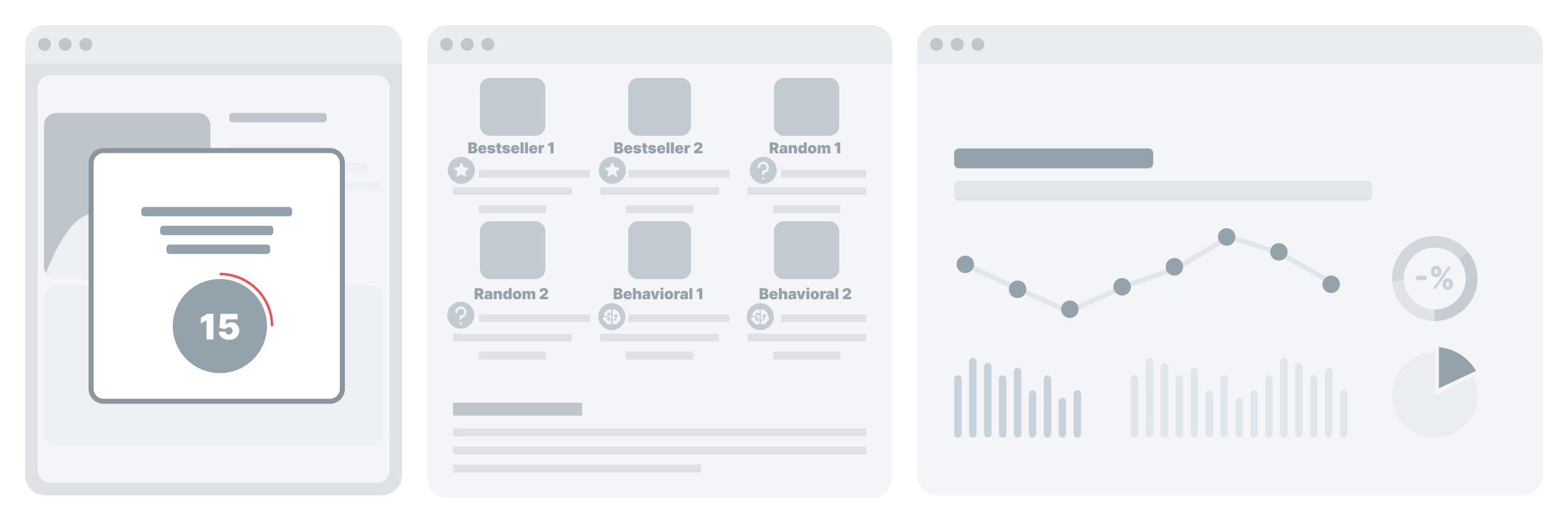}
  \caption{Examples of bright pattern categories: 1) Friction, 2) Transparency, 3) Explainability.}
  \Description{Examples of bright pattern categories: 1) Friction, 2) Transparency, 3) Explainability.}
  \end{centering}
  \label{fig:teaser}
\end{teaserfigure}

\maketitle


\section{Introduction}
As the field of user experience (UX) design has matured, so too has the recognition of the ethical implications of our work. In recent years, designers have come under fire for creating technologies harmful to users, leading to increased scrutiny and criticism from both the design community and lawmakers. While some design practices – such as the use of “dark patterns” to deceive or manipulate users – are clearly unethical, many other practices are more difficult to judge normatively. The rise of artificial intelligence, automation, personalization, and recommendation systems has presented new ethical challenges that require more nuanced solutions. In response, we propose the concept of 'bright patterns' – aspirational design solutions that prioritize ethical considerations over established design conventions, such as those involving reflective design. The ambition of this paper is threefold: to define the term “bright patterns”, to provide examples of such patterns, and to advocate for the adoption of bright patterns through policymaking. By doing so, we hope to foster a culture of responsible design and encourage designers to take a more active role in shaping the policies that govern our work.

\section{Related Work}

Dark patterns are recognized as unethical and increasingly face legal scrutiny; properly conducted user experience design does not lead to designs that overtly deceive users in favor of business interests, but it can result in designs that favor marketable user desires and needs over their long-term goals and well-being. 

\citet{Monteiro2019-ro} argues that designers have a responsibility to ensure the ethical well-being of end-users and to act as the last line of defense against unethical practices. However, the ways in which UX designers engage with ethical issues remain under-researched \cite{Fansher2018-yr}. The term 'dark patterns' originates from software and design patterns. It has various definitions \cite{Brignull2015-zq}. \textbf{We define it as deceptive user interface design functionality that prioritizes business’ objectives over users' well-being and goals.}
 This body of work has led to substantial progress in critical scholarship \cite{Gray2018-mv, Lukoff2021-lz, Mathur2019-do} and legal push-back against manipulative, unethical UX design practices \cite{Berbece2019-vr, Federal_Trade_Commission2022-ra, Wikipedia_contributors2023-sf, Gray2021-sk, Federal_Trade_Commission2021-mr}.

A focus solely on 'dark patterns' and the negative impacts of unethical design practices may be limiting. \citet{Gray2019-kd} present a case study of three UX practitioners and illustrate how designers activate personal and organizational values to navigate complex ethical design considerations. 
While some philosophers reasoned for technologies value neutrality, it is more commonly accepted that technology is value-laden \cite{Franssen2022-ff}. Enabling technology, such as artificial intelligence, personalization, recommender, and automation do present threats to user well-being such as addiction, reliance, privacy, inequality, and bias. These threats are hard to judge normatively out of context \cite{Milano2020-lo, Chellappa2005-df, Bostrom2014-lf}; the morality of technology depends on how the technology is crafted and put to use.  

Designers deal with highly contextual problems regularly. Work on so-called wicked problems is nonlinear, without a definitive, and no \textit{right} solution. The problem-solving process ends when one runs out of resources \cite{Rittel1974-fg}. According to \citet{Sweeting2015-yb} ethical design work is hidden and only implicit;  working on wicked problems is in some ways like working on ethical dilemmas. In these cases, the harms and benefits of technology cannot fully be resolved; \textit{Designers have no way to be right, but also no right to be wrong}. UX designers recognize that many designs cannot easily be distinguished in \textit{bright} or \textit{dark} \cite{Alavi2020-pd}. Designers often find themselves facing complex ethical dilemmas that lack clear solutions, leaving them in a gray area \cite{Sweeting2018-lr}. 

In western User Interface (UI) design communities, design conventions are well established. \textit{Good design} that promotes efficient, usable interaction is documented in plenty of design principles, guidelines, and patterns \cite{Toxboe2023-tp, Apple_Inc2023-ds, Joyce2023-rf, Skrok2023-wu, Tetzlaff1991-op}. Critical and reflective design approaches call us to question these assumptions. Critical design shall foreground ethics of design practice and help to reveal hidden agendas and values by provocation \cite{Bardzell2012-uj}.  Reflective design, build on critical design approaches, uses reflection to uncover and challenge assumptions and biases in design practice, to “bring unconscious aspects of experience to conscious awareness, thereby making them available for conscious choice” \cite{Sengers2005-rf}. Critical reflection shall be applied by both users and designers. Despite its popularity in Human-Computer-Interaction (HCI), critical design is not well adopted in the practicing communities \cite{Bardzell2013-vs}. Applying reflective design does frequently require breaking well established conventions of \textit{good design}, such as usability \cite{Quanjer2014-th}. 

Digital companies prioritize using measurable results, such as key results and performance metrics. Usability and user experience design metrics are used for assessment of design quality and systematically applied in UX teams \cite{Stray2022-nt, Knight2019-gb, Liikkanen2014-mj, Eriksson2021-ba, Schrepp2017-xg}.
UX designers should ensure that human-centered technology is built, typically through user need-based design \cite{Green1990-rl}. The responsibility for ethical design outcomes, though, cannot rest solely on individual designers, as organizational goals often do not align with ethical goals \citep{Berdichevsky1999-bq, Kerssens-van_Drongelen2003-zv, Dourish2019-wp} and designers may act immorally due to self-deception \citep{Tenbrunsel2004-rv}. This misalignment can lead to unintentional or even unwilling ethical violations in the design of digital technology.

While dark patterns are well understood, the literature shows ambiguity around what constitutes good and what ethical design patterns. To our knowledge, no one has systemically considered what the antonym to dark pattern design is. The concept of 'bright patterns' can provide a positive framework, for ethical design. 
It is not yet established or defined, and there also has not been an analysis of existing examples in the field. 
One journal article uses the term bright patterns to describe privacy-friendly nudges \cite{Grasl2021-vx}, one blog post to describe generally good design practices \cite{Coleman2019-aq}, and one master's thesis to describe UI that persuades consumers to make decisions that benefit both themselves and the company \cite{Truong1680425}.  
Further research is needed to fully develop the concept.

\section{Defining Bright Patterns} 

While existing literature has made significant strides in identifying and combatting unethical design practices, UX design can lead to ethically ambiguous design; the concept of 'bright patterns' can provide a proactive and positive frame for ethical design that keeps business and users themselves from inflicting harm.

\textbf{
Bright patterns shall refer to persuasive user interface design functionality that prioritizes users’ well-being and goals over their desires and business' objectives.
}

While the term dark patterns demands an antonym to exist, it is not obvious what it would be; the few examples mentioning 'bright patterns' use it incoherently. To remind ourselves, dark pattern is a term used in user experience (UX) design to describe user interfaces that are intentionally designed to benefit businesses and deceive users into taking actions they would not otherwise take. Design absent of dark patterns, i.e., functional interfaces that do not deceive, would be labeled simply good or effective design. Good design describes functional, aesthetically pleasing products, services, or experiences that meet the needs of their intended users. Poor design, on the other hand, i.e., design that does not work as intended, is sometimes referred to as 'anti-pattern' \cite{Schell2015-kp, Tiangpanich2022-vb, MacDonald2019-kj}. 
Additional confusion may arise as dark patterns are sometimes used to colloquially refer to any type of unethical user interface \cite{Gray2018-mv}, and not just those that deceive and favor business needs over users'. Further, UX designer today is often used ambiguously as a synonym for user interface designer, for example in job listings. 

If user interface designers would correctly use user experience design methods in their work, i.e., user-centered, user-needs-based design, such dark patterns should not come to exist in the first place. Needs-based UX design can still lead to interface design with questionable ethics, when commodifiable needs are prioritized over those that lead to long term flourishing, and well-being of a user. Humans are complex, with diverse desires, needs, and goals that often times are conflicting within themselves. We see in particular trends towards automation, personalization, and recommendation exemplary for such potential conflict. 

Bright patterns, as an antonym to dark patterns, cannot just describe good design practice or the absence of dark patterns. We define bright patterns as user interface elements that promote user behavior aligned with their genuine goals, rather than their immediate desires or businesses' objectives. Notably, these patterns are intentionally designed by businesses to resist short-term gains that would come at the expense of user long-term satisfaction, despite the conflict of interest between the two parties.

\subsection{Examples}
Based on our definition of bright patterns, we have collected examples of bright pattern categories for the themes of friction, transparency, and explainability. We have set up the website \href{https://brightpatterns.org}{brightpatterns.org} for collection and maintenance of bright pattern examples in the wild. This collection is not exhaustive, but demonstrates the characteristics of these patterns and highlights that some businesses are already using them, albeit without explicit acknowledgement. 
 

\begin{itemize}
    \item Slow down: This pattern involves friction in interaction by adding extra steps or barriers to prevent users from engaging in harmful or addictive behaviors. For example, a social media app may require users to confirm their intention before posting a potentially toxic comment, or an app intentionally take longer to open up. \cite{Gruning2023-ou, Porter2019-kr}
    \item Escape hatch: Making it easy for users to leave a situation or cancel a subscription. For example, providing a clear and accessible option to cancel a membership or delete a profile. \cite{Apple_Inc2022-lu, Microsoft2018-df}
    \item Simple Consent: This pattern involves obtaining clear and unambiguous consent from the user before collecting, using, or sharing their personal data. This can involve providing clear explanations of how the data will be used and providing easy-to-use options for opting in or out. \cite{Awwards2021-ab, Kaminska2022-tu, Davis2017-fv}
      \item Honest defaults: This pattern involves setting default options that are in the best interest of the user, rather than the business. For example, a default option to unsubscribe from marketing emails may be provided, rather than requiring users to opt-out. \cite{Toxboe2022-rc}
   \item Nutrition labels: The nutrition label is a design element that provides users with a clear and transparent overview about the content or impact of a particular action, decision, or data. These are becoming common in AI dataset documentation. \cite{Barr2021-ld, Consumer_Reports2021-tg}
    \item Persona profiling:  The app or website transparently informs users of the categories or groups they are being placed into based on their data, allowing them to better understand how they are being profiled and how their experience may differ from other users. \cite{Li2022-qb}
    \item Healthy alternative/5-a-day: The platform suggests higher quality content that is still of interest to the user, but may discourage mindless usage. This pattern respects the user’s well-being and attention, and does not try to exploit their curiosity or boredom into consuming more content than they need or want. or example, a social media app may suggest that users watch educational videos or read informative articles instead of scrolling through endless feeds of memes or gossip. \cite{Kantrowitz2021-zr}
    \item Usage limits: Often implemented in external well-being and child control apps, these are patterns within apps that limit usage time to healthy levels. \cite{BBC_News2021-fe}
    \item Transparent recommender: This pattern involves revealing the logic or criteria behind the recommendations or suggestions that are provided to the user. For example, a streaming service may explain why a certain show or movie is recommended based on the user’s preferences, ratings, or viewing history, or an e-commerce site may disclose how sponsored products are ranked or selected. \cite{Ngo2020-eh,Salesforce_Inc2021-ch, Swayne2021-dw}
    \item Dumb it down: This pattern involves providing users with clear and understandable explanations of how their data is processed or used by an AI system. These explanations can be visual, personalized, and even counterfactual. For example, a credit-scoring system may provide users with a personalized explanation of how it determines their creditworthiness based on their data. \cite{Byrne2019-np, People_AI_Research_team2020-ig}
    \item Data traces: This pattern involves showing users the traces or records of their data that are stored, shared, or accessed by an app or website. For example, a messaging app may show users when their messages are read, forwarded, or deleted by others, or a search engine may show users their search history and explain how it affects their results. \cite{Davies2018-ga, Nissenbaum2023-jx}
    \item Cost transparency: This pattern involves showing users the detailed breakdown of the costs involved in producing and selling a product or service. This pattern respects the user’s curiosity and trust, and does not try to hide or inflate the margins or profits of the vendor. For example, showing users how much the product costs in terms of purchase, marketing, return and shipping. \cite{Airbnb2022-af}
    \item Outside my bubble: This pattern involves exposing users to different perspectives or opinions frequently that challenge their existing beliefs or preferences. For example, a news aggregator may show users articles from diverse sources or a music streaming service may suggest songs from genres that they usually don’t listen to.~\cite{Smith2017-lu, Ovide2021-ny, Adee2016-kv, Peters2023-to}
    \item Contrarian’s companion: This pattern provides critique on consumed and posted content. For example, it shows, views with different political leanings than theirs, and makes users reflect on their opinions. \cite{Feuer2021-km}

\end{itemize}

\subsection{Promoting Bright Patterns through Policy}

Implementing policy-driven change is a difficult task, and it cannot be guaranteed that the policy solutions chosen will be effectively communicated and understood, properly executed, or result in the intended outcomes \cite{Hudson2019-fy}. While ambiguity is part of policy adoption, it can lead to numerous problems in implementation \cite{May2015-gd, Matland1995-zp, Vayrynen2022-sw}. Policy making which relies on ambiguous terms can lead to non-optimal implementations, with businesses potentially mixing deceptive practices into their chosen implementations. Speaking about and defining actual patterns can help to avoid this issue.

A clear definition of 'bright patterns' is necessary for policymakers to start regulating the design of ethical user interfaces that prioritize
user goals and well-being over their desires and business objectives. Defining and recognizing bright patterns can help policymakers establish legal guidelines for user interface design that prioritize these principles over business interests. In some jurisdictions, the regulation of user interface elements, which can be considered as bright patterns, has already begun.

The most well-known example of policy endorsed ethical UI pattern is the cookie consent pop-up in the European Union, which is required by the General Data Protection Regulation (GDPR) \cite{Koch2019-rj}. However, the cookie consent debacle, where companies employed dark patterns to deceive users into accepting privacy settings they would not have chosen otherwise, shows how important it is to design these user interfaces with business buy-in and in user-friendly ways; otherwise, bright patterns risk being ineffective annoyances which businesses circumvent through manipulative practices \cite{Machuletz2020-ct}.

Another recent example, of legislation promoting ethical designs, are screen time limits in China. The government has implemented regulations on the design of user interfaces for online games to restrict the amount of time minors spend playing them \cite{Goh2021-vh}. These regulations require game companies to implement features such as daily play time limits and age-appropriate content.

Moreover, the establishment of clear definitions and guidelines for bright patterns can lead to a more honest and self-conscious UX design community. It is desirable that bright patterns are invented and designed through teams of UX designers applying user-centered design practices, with the intrinsic motivation of designers to build long-term trust and reputation. Governmental agencies rarely have the in-house power to do so. Leading companies need to set benevolent examples, as advocating for bright patterns from internal motivations will motivate more businesses to set good practice examples.

To effectively regulate and regularize bright patterns, policymakers and the UX design community must collaborate. UX designers can offer expertise and insight into how to design interfaces that prioritize user needs, while policymakers can provide guidance on how to create legal frameworks that promote fair and ethical practices in UX design \cite{W3C_Web_Accessibility_Initiative_WAI2018-zo}.

Overall, while there are currently only a few specific user interface patterns that are regulated by law, the increasing interest in regulating other aspects of user interface design, particularly regarding dark patterns, is a great starting point for policymaking, UX design, and HCI researcher communities to come together to innovate policy which does not just prevent overt harm but also promotes good.

\section{Discussion} 
The emergence of dark patterns in user interface design raises significant ethical concerns. Businesses use these deceptive techniques to increase profits, often at the expense of users' autonomy, trust, and well-being. With the rise of regulations against dark patterns, it is now the time to define the opposite concept of bright patterns.
Bright patterns, as we define them, use persuasion rather than deception to prioritize users' goals and well-being over their desires and business objectives. Bright patterns can promote trust, foster positive relationships between businesses and their users, and contribute to user autonomy, well-being, and overall experience.

Focusing on patterns, as in standardized design solutions, has several advantages for policy making over regulating design processes. Design processes can be complex and varied, making them difficult to regulate with a one-size-fits-all policy. A policy aimed at regulating design processes might be too broad and could lead to unintended consequences, such as stifling innovation. Regulating exact design patterns can provide clearer and more actionable guidance to designers, be more effective in promoting well-being to users, and be more flexible in responding to changing design practices.

As regulating bright patterns on their own can be a challenging task for policymakers, collaboration between sectors and fields is compulsory. The professional UX design community and the academic HCI community can contribute to this effort by providing insights on best practices, user-centered design, and ethics. Policymakers can use this information to establish guidelines and regulations that promote ethical user interface design. We hope that our short paper encourages future work on this necessary thread of work.

We acknowledge that the terms dark pattern and bright pattern build on dualistic views, equating bright with good and dark with evil \cite{Banerjee2012-gs}. Alternative terms good design, bad design, ethical design, and unethical design, are all describing different concepts. Harry Brignull is credited with coining the term 'dark patterns' \cite{Brignull2015-zq}; he has since switched to using the term 'deceptive design', citing concerns about the precision and inclusivity of the term. The terms 'honest design' and 'transparent design' are close antonyms to dark patterns, but they do not adequately capture the ethical tension and prioritization of user goals over user needs and business interests that bright patterns embody. 
Furthermore, we find the term 'design' to be ambiguous, as it can refer to the design process, intended design, and design outcome. Therefore, we suggest when referring to ethical design practices that go beyond established moral standards, one may refer to \emph{benevolent design}. 


\begin{acks}
This work originated from a class in philosophical and analytical approaches to societal implications of digital technologies. I thank Professor Helen Nissenbaum for her mentorship, and classmate Amritansh Kwatra for his help in outlining this thread of work.  
\end{acks}

\bibliographystyle{ACM-Reference-Format}
\bibliography{Sandhaus_6210, Sandhaus_manual, bright-patterns-examples}


\end{document}